\documentclass[conference]{IEEEtran}

\usepackage{graphicx,subcaption}
\usepackage{color}
\usepackage{enumitem}
\usepackage{amsmath}
\usepackage{amsthm}
\usepackage{wrapfig}
\usepackage{xspace}
\usepackage{algcompatible}
\usepackage{algorithm}
\usepackage{algpseudocode}
\usepackage{listings}
\usepackage{verbatim}
\usepackage{multicol}
\usepackage{float}
\usepackage[utf8]{inputenc}             
\usepackage[T1]{fontenc}                
\usepackage{hyperref}
\usepackage{booktabs}                  
\usepackage{amsfonts}                  
\usepackage{nicefrac}                  
\usepackage{microtype}                 
\usepackage{listings}
\usepackage{xcolor}

\usepackage{pifont}
\usepackage{subfiles}
\usepackage{multirow}
\usepackage{colortbl}
\usepackage[thicklines]{cancel}

\graphicspath{ {figs/} }

\algnewcommand{\IfThenElse}[3]{
  \State \algorithmicif\ #1\ \algorithmicthen\ #2\ \algorithmicelse\ #3}
\algnewcommand{\IfThen}[2]{
  \State \algorithmicif\ #1\ \algorithmicthen\ #2}  
\algnewcommand{\IfThenNoS}[2]{
 \algorithmicif\ #1\ \algorithmicthen\ #2}  
\algnewcommand{\IfNoS}[2]{
 \algorithmicif\ #1\ \algorithmicthen\ #2}  
\begin{document}
\newcommand{\punt}[1]{}
\newcommand{\cmnt}[1]{}

\definecolor{xxxcolor}{rgb}{0.8,0,0}
\newcommand{\XXX}[1]{{\color{xxxcolor} XXX: #1}\xspace}

\newcommand{\func}[1]{\texttt{#1}}
\newcommand{\var}[1]{\texttt{#1}}

\newcommand{\nosplit}{\linebreak}

\def\nohyphens{\hyphenpenalty=10000\exhyphenpenalty=10000}

\newcommand{\tilda}{\symbol{126}}

\newcommand{\ang}[1]{\langle #1 \rangle}
\newcommand{\Ang}[1]{\Big\langle #1 \Big\rangle}
\newcommand{\ceil}[1]{\lceil #1 \rceil}
\newcommand{\floor}[1]{\lfloor #1 \rfloor}

\newtheorem{theorem}{Theorem}
\newtheorem{lemma}[theorem]{Lemma}
\newtheorem{corollary}[theorem]{Corollary}
\newtheorem{proposition}[theorem]{Proposition}
\newtheorem{property}[theorem]{Property}
\newtheorem{claim}[theorem]{Claim}
\newtheorem{definition}{Definition}
\newtheorem{guideline}{Rule}
\newtheorem{requirement}[theorem]{Requirement}
\newcounter{history}
\newcommand{\hist}[1]{\refstepcounter{history} {#1}}

\newcommand{\ajay}[1]{\textcolor{red}{[[[Ajay: #1]]]}}


\newtheorem{acknowledgement}[theorem]{Acknowledgement}
\newtheorem{observation}[theorem]{Observation}
\newtheorem{assumption}[theorem]{Assumption}

\newcommand{\chapref}[1]{Chapter~\ref{chap:#1}}
\newcommand{\secref}[1]{Section~\ref{sec:#1}}
\newcommand{\figref}[1]{Figure~\ref{fig:#1}}
\newcommand{\tabref}[1]{Table~\ref{tab:#1}}
\newcommand{\stref}[1]{step~\ref{step:#1}}
\newcommand{\thmref}[1]{Theorem~\ref{thm:#1}}
\newcommand{\lemref}[1]{Lemma~\ref{lem:#1}}
\newcommand{\insref}[1]{line~\ref{ins:#1}}
\newcommand{\corref}[1]{Corollary~\ref{cor:#1}}
\newcommand{\axmref}[1]{Proposition~\ref{axm:#1}}
\newcommand{\defref}[1]{Definition~\ref{def:#1}}
\newcommand{\eqnref}[1]{Eqn(\ref{eq:#1})}
\newcommand{\eqvref}[1]{Equivalence~(\ref{eqv:#1})}
\newcommand{\ineqref}[1]{Inequality~(\ref{ineq:#1})}
\newcommand{\exref}[1]{Example~\ref{ex:#1}}
\newcommand{\propref}[1]{Property~\ref{prop:#1}}
\newcommand{\clmref}[1]{Claim~\ref{clm:#1}}
\newcommand{\obsref}[1]{Observation~\ref{obs:#1}}
\newcommand{\asmref}[1]{Assumption~\ref{asm:#1}}
\newcommand{\thref}[1]{Thread~\ref{th:#1}}
\newcommand{\trnref}[1]{Transaction~\ref{trn:#1}}
\newcommand{\lstref}[1]{listing~\ref{lst:#1}}
\newcommand{\Lstref}[1]{Listing~\ref{lst:#1}}

\newcommand{\subsecref}[1]{SubSection{\ref{subsec:#1}}}

\newcommand{\histref}[1]{\ref{hist:#1}}

\newcommand{\apnref}[1]{Appendix~\ref{apn:#1}}
\newcommand{\invref}[1]{Invariant~\ref{inv:#1}}

\newcommand{\Chapref}[1]{Chapter~\ref{chap:#1}}
\newcommand{\Secref}[1]{Section~\ref{sec:#1}}
\newcommand{\Figref}[1]{Figure~\ref{fig:#1}}
\newcommand{\Tabref}[1]{Table~\ref{tab:#1}}
\newcommand{\Stref}[1]{Step~\ref{step:#1}}
\newcommand{\Thmref}[1]{Theorem~\ref{thm:#1}}
\newcommand{\Lemref}[1]{Lemma~\ref{lem:#1}}
\newcommand{\Corref}[1]{Corollary~\ref{cor:#1}}
\newcommand{\Axmref}[1]{Proposition~\ref{axm:#1}}
\newcommand{\Defref}[1]{Definition~\ref{def:#1}}
\newcommand{\Eqref}[1]{eq(\ref{eq:#1})}
\newcommand{\Eqvref}[1]{Equivalence~(\ref{eqv:#1})}
\newcommand{\Ineqref}[1]{Inequality~(\ref{ineq:#1})}
\newcommand{\Exref}[1]{Example~\ref{ex:#1}}
\newcommand{\Propref}[1]{Property~\ref{prop:#1}}
\newcommand{\Obsref}[1]{Observation~\ref{obs:#1}}
\newcommand{\Asmref}[1]{Assumption~\ref{asm:#1}}
\newcommand{\reqref}[1]{Requirement~\ref{req:#1}}

\newcommand{\Lineref}[1]{Line~\ref{lin:#1}}
\newcommand{\lineref}[1]{line~\ref{lin:#1}}
\newcommand{\algoref}[1]{Algorithm~\ref{algo:#1}}
\newcommand{\Algoref}[1]{{\sf Algorithm$_{\ref{algo:#1}}$}}

\newcommand{\Apnref}[1]{Section~\ref{apn:#1}}
\newcommand{\Invref}[1]{Invariant~\ref{inv:#1}}
\newcommand{\Confref}[1]{Conflict~\ref{conf:#1}}
\newcommand{\ruleref}[1]{Rule~\ref{rul:#1}}

\newcommand{\theqed}{$\Box$}
\newcommand{\nsqed}{\hspace*{\fill} \theqed}

\renewcommand{\thefootnote}{\alph{footnote}}
\newcommand{\ignore}[1]{}
\newcommand{\myparagraph}[1]{\noindent\textbf{#1}}

%

\newcommand{\lastup} {lastUpdt}
\newcommand{\lupdt}[2] {#2.lastUpdt(#1)}
\newcommand{\fkmth}[3] {#3.firstKeyMth(#1, #2)}

%
\newcommand{\nz}{\emph{restartable}\xspace}
\newcommand{\cas}[3] {CAS(#1, #2, #3)}

\newcommand{\qp}{\emph{quiescent-phase}\xspace}
\newcommand{\rdp}{\emph{$\Phi_{read}$}\xspace}
\newcommand{\wtp}{\emph{$\Phi_{write}$}\xspace}

\newcommand{\rd}{\emph{reader}\xspace}
\newcommand{\wt}{\emph{writer}\xspace}
\newcommand{\rl}{\emph{reclaimer}\xspace}
\newcommand{\rrc}{\emph{reader-reclaimer}\xspace}
\newcommand{\wrc}{\emph{writer-reclaimer}\xspace}
\newcommand{\nbr}{\emph{NBR}\xspace}
\newcommand{\nbrp}{\emph{NBR+}\xspace}
\newcommand{\knbr}{\emph{NBR}\xspace}
\newcommand{\ds}{\emph{data-structure}\xspace}
\newcommand{\rb}{\emph{retireBag}\xspace}
\newcommand{\lb}{\emph{limboBag}\xspace}
\newcommand{\hw}{\emph{HiWatermark}\xspace}
\newcommand{\lw}{\emph{LoWatermark}\xspace}
\newcommand{\smr}{\emph{safe memory reclamation}\xspace}

\newcommand{\tid}[1]{$T_#1$}
\newcommand{\pred}{\texttt{pred}\xspace}
\newcommand{\curr}{\texttt{curr}\xspace}
\newcommand{\suc}{\texttt{succ}\xspace}

\newcommand{\rcu}{\emph{RCU}\xspace}
\newcommand{\qsbr}{\emph{QSBR}\xspace}
\newcommand{\debra}{\emph{DEBRA}\xspace}
\newcommand{\ibr}{\emph{IBR}\xspace}
\newcommand{\hp}{\emph{HP}\xspace}
\newcommand{\he}{\emph{HE}\xspace}
\newcommand{\ebr}{\emph{EBR}\xspace}
\newcommand{\rgp}{\emph{RGP}\xspace}
\newcommand{\rev}{\emph{reclamation event}\xspace}

%
\newcommand{\crdp}[1]{\texttt{cread}(#1)}
\newcommand{\cwrp}[2]{\texttt{cwrite}(#1, #2)}
\newcommand{\cwr}{\texttt{cwrite}\xspace}
\newcommand{\crd}{\texttt{cread}\xspace}
\newcommand{\utag}[1]{\texttt{untagOne} #1}
\newcommand{\utagp}[1]{\texttt{untagOne}(#1)}
\newcommand{\utagall}{\texttt{untagAll}\xspace}
\newcommand{\hbrbit}{\textit{accessRevokedBit}\xspace}
\newcommand{\hbrset}{\textit{tagSet}\xspace}
\newcommand{\head}{\textit{head}\xspace}
\newcommand{\ca}{\textit{Conditional Access}\xspace}
\newcommand{\uaf}{\textit{use-after-free}\xspace}

\newcommand{\ajedit}[1]{\textcolor{red}{#1}}
 
\title{Efficient Hardware Primitives for Immediate Memory Reclamation in Optimistic Data Structures
}
\author{\IEEEauthorblockN{Ajay Singh}
\IEEEauthorblockA{\textit{University of Waterloo} \\
ajay.singh1@uwaterloo.ca}
\and
\IEEEauthorblockN{Trevor Brown}
\IEEEauthorblockA{\textit{University of Waterloo} \\
trevor.brown@uwaterloo.ca}
\and
\IEEEauthorblockN{Michael Spear}
\IEEEauthorblockA{\textit{Lehigh University} \\
spear@lehigh.edu}
}

\maketitle

\begin{abstract}
Safe memory reclamation (SMR) algorithms are crucial for preventing use-after-free errors in optimistic data structures.
SMR algorithms typically delay reclamation for safety and reclaim objects in batches for efficiency.
It is difficult to strike a balance between performance and space efficiency.
Small batch sizes and frequent reclamation attempts lead to high overhead, while freeing large batches can lead to long program interruptions and high memory footprints.
An ideal SMR algorithm would forgo batching, and reclaim memory immediately, without suffering high reclamation overheads.

To this end, we propose \textit{Conditional Access}: a set of hardware instructions that offer immediate reclamation and low overhead in optimistic data structures.
\textit{Conditional Access} 
harnesses cache coherence to enable threads to efficiently \textit{detect} potential use-after-free errors without explicit shared memory communication, and without introducing additional coherence traffic.

We implement and evaluate \ca in Graphite, a multicore simulator. 
Our experiments show that \textit{Conditional Access} can rival the performance of highly optimized and carefully tuned SMR algorithms while simultaneously allowing immediate reclamation.
This results in concurrent data structures with similar memory footprints to their sequential counterparts.
\end{abstract}

\begin{IEEEkeywords}
Safe Memory Reclamation, Optimistic Data Structures, Shared Memory Data Structures
\end{IEEEkeywords}

\lstdefinestyle{ajstyle}{ %
	language=C++,
	numbers=left,                    
	morekeywords={*, procedure,function},   
	breaklines=true,                 
	frame=single,
belowcaptionskip=1\baselineskip,
showstringspaces=false,
basicstyle=\scriptsize\ttfamily,
keywordstyle=\bfseries\color{green!40!black},
commentstyle=\itshape\color{purple!40!black},
identifierstyle=\color{blue},
stringstyle=\color{orange},	
}

\lstdefinestyle{ajstyleds}{ %
	language=C++,
	morekeywords={*, startOp,...},   
	breaklines=true,                 
	frame=single,
belowcaptionskip=1\baselineskip,
xleftmargin=\parindent,	
showstringspaces=false,
basicstyle=\scriptsize\ttfamily,
keywordstyle=\bfseries\color{green!40!black},
commentstyle=\itshape\color{purple!40!black},
identifierstyle=\color{blue},
stringstyle=\color{orange},	
numberstyle=\scriptsize,
}
\lstset{escapechar=|,style=ajstyle}

\section{Introduction}
\label{sec:intro}
Current \smr(SMR)~\cite{brown2015reclaiming, michael2004hazard, alistarh2014stacktrack, alistarh2017forkscan, alistarh2018threadscan, balmau2016fast, cohen2018every, nikolaev2019hyaline, wen2018interval, detlefs2002lock, hart2007performance, nikolaev2020universal, singh21nbr} algorithms used in many optimistic data structures \textit{delay reclamation} and free nodes in batches to trade-off space in favor of high performance and safety.
When the batches are too small, the data structure's \textit{throughput} suffers due to more overhead from frequent reclamation. On the other hand, when the batches are too large, 
though the reclamation overhead is amortized due to reduced frequency of reclamation, 
the occasional freeing of large batches causes long program interruptions and  dramatically increases \textit{tail latency} for data structure operations.%

Larger batch sizes also increase the \textit{memory footprint} of applications, which makes memory utilization and allocation challenging in virtualized environments~\cite{vmwareunderstanding}.
For example, increased memory footprints of virtual machines (VMs) or processes due to large batch sizes preclude the host machine from taking advantage of \textit{Memory Overcommitment} where the available dynamic memory could otherwise be shared amongst multiple VM instances (or other processes).

Besides requiring programmers to find an acceptable batch size (i.e., \textit{reclamation frequency}), most fast epoch-based SMR algorithms \cite{brown2015reclaiming, nikolaev2019hyaline, wen2018interval, nikolaev2021crystalline} also have to determine an optimal increment frequency of a global timestamp (sometimes referred to as \textit{epoch frequency}). The values of these parameters in tandem influence the time and space efficiency of SMR algorithms.
Choosing an optimal value for these parameters can be quite challenging since they vary depending on the type of data structure, workload, and machine characteristics.

\ignore{
In lock based linked data structures safely reclaiming memory is straightforward.
For example, in a linked list, a thread can delete a node by first locking it and its predecessor, then unlinking the node and immediately freeing it to the operating system, and finally unlocking the predecessor.
This is safe, because while the predecessor is locked, no other thread can have a pointer to the node being deleted.

However, modern concurrent linked data structures improve performance by allowing threads to perform \textit{optimistic reads} of the contents of nodes \textit{without} locking them.
Since one can never know if a node is being read, nodes cannot be freed immediately, as doing so might cause a reader to crash.
In such data structures, one typically uses a \textit{safe memory reclamation} (SMR) algorithm, which \textit{delays} reclamation of an unlinked node until it can guarantee that no thread has a pointer to the node~\cite{brown2015reclaiming, michael2004hazard, alistarh2014stacktrack, alistarh2017forkscan, alistarh2018threadscan, balmau2016fast, cohen2018every, nikolaev2019hyaline, wen2018interval, detlefs2002lock, hart2007performance, nikolaev2020universal, singh21nbr}.
Checking whether any thread has a pointer to a particular unlinked node can be expensive, so unlinked nodes are often collected in batches to amortize the overhead of such checks.
This leads to a key tension in SMR algorithms: increasing the batch size amortizes reclamation overheads, but it also delays reclamation of nodes for longer, increasing memory consumption and potentially worsening cache locality. 
\textcolor{blue}{This could result in slowdown due to negative interactions with the underlying memory allocator as shown in \cite{singh21nbr}.}
}

In this work, we turn the traditional SMR paradigm on its head.
Whereas SMR algorithms usually ensure that reclaimers delay reclamation until a node can no longer be accessed by readers, we allow reclaimers to free immediately, and put the onus on readers to avoid inconsistency.
Inspired by the recent \textit{Memory Tagging} proposal of Alistarh et~al.~\cite{alistarh2020memory}, we propose a new hardware mechanism called \textit{Conditional Access} to allow readers to efficiently determine whether a node they are trying to access has been freed.
It allows a thread to \textit{conditionally read} (\texttt{cread}) a new location \textit{only if} a set of programmer defined \textit{tagged} locations
have not changed since they were previously read.
Similarly, threads can conditionally write (\texttt{cwrite}) a tagged location by \texttt{validating} that no tagged location has changed.
Locations are tagged by invoking \texttt{cread}, and are manually untagged by invoking \texttt{untagOne} or \texttt{untagAll}.

\textit{Conditional Access} is ideal for implementing data structures for which 
one can prove a read is safe if a small set of previously read locations have not changed since they were last read.
For example, in a linked list that sets a \textit{marked bit} in a node before deleting it, if a thread reads the next pointer of an unmarked node, and at some later time its marked bit and next pointer have not been changed, then it is still safe to dereference its next pointer.
In such a data structure, once a node is unlinked and marked, it can \textit{immediately} be freed, since doing so will merely cause subsequent \texttt{cread}s or \texttt{cwrite}s on the node to fail, triggering a \textit{restart} (an approach common to many popular SMRs~\cite{michael2004hazard, cohen2018every, wen2018interval, cohen2015automatic}).

\textit{Conditional Access} can be thought of as a generalization of load-link/store-conditional (LL-SC) where the load is also conditional, and the store can depend on many loads.
Whereas an LL effectively tags a location, and an SC untags the location,
in \textit{Conditional Access}, locations are not automatically untagged when they are written.
So, multiple \texttt{cread}s and \texttt{cwrite}s can be performed on the same set (or a dynamically changing set) of tagged locations.

\textit{Conditional Access} also has some similarities to a restricted form of transactional memory.
However, whereas hardware transactional memory (HTM) is increasingly being disabled due to security concerns, we believe \textit{Conditional Access} can be implemented more securely.
For example, since a thread becomes aware of concurrent updates to its tagged nodes only when it performs a \texttt{cread} or \texttt{cwrite} and then checks a status register, we can avoid some timing attacks that are made possible by the immediacy of aborts (as a result of conflicting access by the other threads) in current HTM implementations.

Much of the information needed to efficiently implement \textit{Conditional Access} is already present in modern cache coherence protocols.
We propose a simple extension where tagging is implemented at the L1 cache level without requiring changes to the coherence protocol. 
At a high level, each L1 cache line has an associated \textit{tag} (a single bit). 
This \textit{tag} is set by a \texttt{cread} on any location in that cache line, and unset by an \texttt{untagOne} on a location in that cache line or an \texttt{untagAll} instruction.
%
Each core tracks invalidations of its own tagged locations. 
(In SMT architectures, where $k$ hyperthreads share a core, each hyperthread tracks invalidations of its own tagged locations).

While a detailed implementation at the microarchitectural level is beyond the scope of this paper, the extensions we require to the cache, and between the cache and processor pipeline, are a strict subset of those needed to implement HTM. This strongly suggests that \ca implementations can be practical and efficient.

\ca enables memory footprints similar to those of sequential data structures.
This is desirable in modern data centers, to save costs related to memory over-allocation and to facilitate Memory Overcommitment~\cite{vmwareunderstanding}. 
Further, immediate reclamation can help in avoiding exploits that use the extended lifetime of unlinked objects in delayed reclamation algorithms to leak private data.
It also has the potential to prevent denial of service attacks in which threads induce a schedule that causes batches of unreclaimed memory to grow unboundedly, leading to out-of-memory errors.
Such attacks have been reported in RCU implementations in the Linux kernel~\cite{mckenney2006extending}.


This paper makes the following contributions.
(A) We introduce \ca, a set of hardware instructions that enable efficient immediate memory reclamation. (B) We prototype \ca on an open source multicore simulator, Graphite. (C) We implement a benchmark comprised of multiple state-of-the-art memory reclamation techniques and data structures. 
(D) We show how \ca can be used to avoid \uaf errors for several data structure design patterns, including optimistic two-phased locking.

The remainder of the paper is organized as follows.
\secref{hbrspec} explains the core idea of \ca and the semantics of the proposed instructions.
\secref{hwimpl} sketches a straightforward hardware implementation of \ca.
\secref{applyin-to-ds} discusses how \ca can be used with optimistic data structures, using a stack and a lazy linked list as examples, and briefly discusses correctness.
\secref{exp} benchmarks \ca, illustrating its efficiency and low memory usage.
Related work appears in \secref{related}, followed by a conclusion in \secref{conclusion}. 

\section{Conditional Access}
\label{sec:hbrspec}
\subsection{Key Idea}
A \uaf error can be considered a special case of a read-write data race, where a shared memory location is accessed after it has been freed by a different thread.
In modern systems with coherent caches, \uaf errors are always preceded by events in the cache coherence protocol.
Consider a traditional MESI protocol: To store a value at a location \texttt{X} that is currently in the shared state, a core $C$ first invalidates copies of \texttt{X} at other cores by sending \textit{invalidation} messages to all other cores.
Upon receiving such a message, a core \textit{invalidates} its copy of that location and responds with an acknowledgement message.
Once $C$ has received acknowledgements from all other cores, it has exclusive access to \texttt{X}.
A thread that reads \texttt{X} after it is freed will respond to such invalidation messages before reading \texttt{X}.
This reveals that, at the level of the coherence protocol, readers are aware that the memory location they are trying to access may have been concurrently modified. 
Moreover, a subsequent read of \texttt{X} must begin with a cache miss\textemdash ~an avoidable overhead if the information about concurrent modification could be harnessed.

Read-write data races and \uaf errors are 
indistinguishable at the architectural level, which makes it difficult to identify \uaf errors by looking solely at events in the cache coherence protocol. 
If we are willing to accept false positives, we can interpret each invalidation message as a sign of a possible \uaf error.
To that end, \ca monitors the exchange of messages at the architecture level between an updating and a reading thread, and exposes these interactions to the program through specialized instructions to enable safe memory reclamation.

We expect the programmer to know which memory accesses \textit{might} result in \uaf errors, and we require the programmer to use our new instructions to perform these accesses.
The hardware can then \emph{tag} the corresponding cache lines, indicating to the hardware that \textit{invalidation} of such a cache line is an event of interest.
Subsequent loads and stores of a tagged location only complete if the location has not been \textit{invalidated} since the location was tagged.
Note that in order for this technique to work, reclaimers must do a store on some tagged location before freeing a node, so they can be sure to trigger a cache event that revokes other threads' access to that node.
The reclaimer can then immediately free the node: Any thread that has tagged this node before it was freed, and subsequently tries to access the tagged node, will observe that the corresponding cache line has been \textit{invalidated}, and will not perform the access.

Since the memory accesses are \textit{conditioned} upon whether a memory location has been invalidated since it was previously tagged, we call our technique \ca.

\subsection{New State and Instructions}

\noindent\textbf{Additional Storage}: (A) Each core \textit{tags} an address for which it intends to monitor invalidation requests. 
Abstractly, the set of tagged addresses can be represented by a \hbrset.
(B) Additionally, each core also maintains an \hbrbit, which is initially clear, and is set when its access is revoked for any of the addresses in its \hbrset. 
For brevity, in this section, we assume that \hbrset's capacity is not bounded. 
Efficiently approximating this set is the subject of
Section~\ref{sec:hwimpl}.

\medskip
\noindent\textbf{Remote Events}: For each entry in a core $C$'s \hbrset, the hardware is required to detect whether any other core has invalidated that cache line since $C$ tagged it.
If another core invalidates this cache line, 
the hardware must set $C$'s \hbrbit.

Having described the \hbrset and \hbrbit, we now describe the proposed instructions:
\\
\noindent \textbf{(1) \func{\crd addr, dest}:}
Similar to a \texttt{load} instruction, \crd updates register \texttt{dest} with the value at the address in register \texttt{addr}, but with two key differences: \textit{tagging} and \textit{conditional access}\footnote{For simplicity of
  presentation, we do not parameterize \crd by the number of bytes to read
  from memory, or consider different addressing modes.  In a practical
  system, several opcodes will be needed for these purposes.}. 
%
More specifically, \crd atomically
checks if \texttt{addr} is in \hbrset, and if not, adds it to \hbrset. 
It also checks if \hbrbit is set, and if so, skips the load, and updates some other processor state, such as a flag register, to indicate that there may have been a \uaf error.
In this case, we say the \crd{} has \textit{failed}.
Otherwise, it loads the value at \texttt{addr} into \texttt{dest}, indicating that the memory access was safe.
In this case, we say the \crd{} has \textit{succeeded}. 
\\
\noindent
\textbf{(2) \func{\cwr addr, v}:}
Unlike \crd, 
\cwr does not update \hbrset.
Atomically: 
\cwr checks if the \hbrbit is set or \texttt{addr} is not in the \hbrset, in which case the store is skipped and a processor flag is set to indicate that the \cwr has \textit{failed} (suggesting there may have been a \uaf error).
Otherwise, it stores \texttt{v} at \texttt{addr}, and we say the \cwr has \textit{succeeded}.

It is worth discussing here why \cwr fails when it executes on an \texttt{addr} which is not in the \hbrset.
This design decision rules out uses where programmers may invoke \cwr before invoking a \crd (or in other words before first tagging a location). This helps to avoid tagging during a \cwr, which could incur significant delays if the access misses in the L1 Cache, making it easier to avoid tricky time-of-check to time-of-use (TOCTOU) issues. In particular we would prefer to avoid scenarios where a \cwr misses in the L1, waits for the data, and takes exclusive ownership of the line, only to discover that the \hbrbit has been set during the wait, thus eventually failing the \cwr. By requiring \crd to be performed first, we move the high latency parts of this operation into a shared mode access, potentially reducing invalidations and coherence traffic.
\\
\noindent
\textbf{(3) \func{\utag{addr}}:}
The \utag{} instruction does not access memory.
Its purpose is to allow the programmer to remove an address from the \hbrset.
If \texttt{addr} is not in \hbrset, \utag{} has no effect. 
Once an address is removed from a core's \hbrset, subsequent remote invalidations of the address will \textit{not} set the core's \hbrbit.
\\
\noindent
\textbf{(4) \func{\utagall}:}
\utagall 
%
clears the \hbrset and unsets the value of \hbrbit.
It is intended to be used in two cases: (1) when a \crd{} or \cwr{} fails, at which point a data structure operation will need to be retried; and (2) before returning from a successful data structure operation.

Note, for SMT architectures with multiple hardware threads \textit{additional storage} and \textit{remote events} are required per hardware thread, instead of per core.
\section{Implementation}
\label{sec:hwimpl}
\ca can be implemented by a straightforward extension of existing caches, such that modifications are only introduced between a processor and its primary cache, e.g., the L1 data cache.
Based on our prototyping on a multicore simulator, we believe these changes are a strict subset of those required to implement HTM, which implies \ca is practical and efficient to implement.

Implementing \ca requires realizing the \hbrset and \hbrbit.
The proposed instructions use those data structures to track relevant invalidation messages, which are generated by the underlying cache coherence protocol. 

(A)~The \hbrset can be approximated by adding one \texttt{tag} bit to each cache line of a core's L1 data cache. 
This is similar to how hardware transactional memory approximates its read and write sets.
(B) The \hbrbit, which tracks the invalidations of the addresses in a thread's \hbrset, requires adding one bit for each core.
One way the \hbrbit could be implemented is by adding it to the condition code or flag registers of the host architecture (e.g., EFLAGS on x86).

Note, in SMT architectures, where k hyperthreads share a core, each hyperthread will track which of its cache lines are tagged and tracks invalidations of its tagged locations. For instance, on a 2-way SMT architecture, two \emph{tag} bits and two \hbrbit{s}, one for each hardware thread, will be required.

Given these changes, we can now harness the cache coherence protocol to detect unsafe accesses.
When a \crd adds an address to the \hbrset, it loads that line into the
cache and sets the \textit{tag bit} for that line. 
%
There are two ways in which the line can subsequently depart the cache:
remote invalidation or a local associativity conflict.
In either case, the cache must notify the hardware thread that its \hbrbit
must be set, so that its subsequent \crd{} or \cwr{} will fail. 
For remote invalidations, doing so must be atomic with acknowledging the remote request.
For associativity conflicts, doing so must be atomic with fetching new data
from the memory hierarchy.
The atomicity requirements for \utag and \utagall are simpler: they cannot
be reordered with respect to loads and stores by the same hardware thread.
Furthermore, \utagall must clear the \hbrbit for future operations.

Besides the aforementioned two ways, In SMT architectures, a thread's \hbrbit can be set upon a write to a shared cache line by another thread (in case of hyperthreading) or on a context switch.
Setting the bit on a context switch is more straightforward to implement since it enables the Operating system to avoid keeping track of invalidations on behalf of switched-out thread. 
These properties provide a foundation for the \ca to be used in multiuser systems.

Intuitively, tagging in \ca facilitates a kind of local protection of shared memory locations that does not trigger any additional coherence traffic. This is contrary to popular paradigms like hazard pointers~\cite{michael2004hazard} or other reservation-based~\cite{wen2018interval} techniques, which always trigger global cache traffic between threads.

The aforementioned way of implementing \hbrset means that the \hbrset size is bounded by the associativity of the cache and therefore \hbrset could overflow.
This would lead to eviction of tagged addresses (in \hbrset), causing \hbrbit to be set.
This, in turn, could lead to spurious failures of subsequent \crd{s} or \cwr{s} which could stall progress.
However, in practice it is not an issue because in most cases the \hbrset is small.
Our experiments (\secref{exp}) show associativity does not have any significant impact on progress for the workloads we consider.

\section{Using Conditional Access with Optimistic Data Structures}
\label{sec:applyin-to-ds}
In this section we discuss how \ca can be used
to achieve safe memory reclamation with optimistic data structures such as lists~\cite{heller2005lazy} and external binary search trees~\cite{ellen2010non}.
Operations of many such data structures have a search phase consisting of multiple reads, wherein a thread continuously traverses the next fields of nodes until it has visited a set of nodes it is interested in, where the operation eventually takes effect.
After reaching the nodes of interest, the operation may perform zero 
or more writes.
For example, in a linked list, a thread might traverse multiple links to find a predecessor and current node where an operation should take effect.

For ease of exposition we assume that each node fits in a single cache line, and a cache line contains only one node.
Thus, adding a node to a core's \hbrset implies adding a cache line containing the node to the core's \hbrset.
We start by stating the following high level directives required for all data structures to be able to correctly use \ca.


(DI) \textbf{Replace and Analyse:} 
\textit{Replace}: all read/write accesses to nodes that can be freed should be substituted by the corresponding \crd{} and \cwr{} instructions. This enables \ca to tag a node and monitor it for concurrent modification and notify programmers by updating a flag register.
\textit{Analyze}: If a \crd{} or \cwr{} fails, the operation should immediately \utagall and retry. A failed instruction implies a node could have been concurrently freed, therefore any future access will not be safe.

(DII) \textbf{Validate Reachablility:}
A node is tagged when it is first \crd{}. 
In order to ensure that the tagged node is valid, it should be verified it was reachable in the data structure after the fact.

We now demonstrate how these directives can be applied to use \ca in different classes of optimistic data structures. Depending upon the data structures, DI could be partially relaxed, as we will see in the example of a lazy list or DII may not be needed as we will see in the example of a lock free stack. The lazy list requires some more rules which are detailed in the \secref{ll}.

\subsection{In Data Structures with Single Writes}
Data structures with a single write in their update phase include some list based stacks~\cite{treiber1986systems} and queues~\cite{michael1996simple}, both of which we have implemented.
For the purpose of illustration, we will consider a list-based unbounded lock-free stack. In such a stack, a \func{push} operation involves reading a \texttt{top} pointer, allocating a new node for a key value to be pushed, and then doing a Compare-and-Swap (CAS)
to set the node as new \texttt{top}. Likewise, a \func{pop} operation consists of reading the \texttt{top}, and then atomically setting the \texttt{top} to its next node. After a \func{pop}, the unlinked node cannot be freed if a concurrent thread might still access it.


The original operations of the stack could be upgraded to enable \ca by simply replacing every read with \crd{} and the CAS with \cwr{} (DI). Then the \func{pop} operation could immediately free the unlinked node as shown in  \algoref{stack}.
Note, in our pseudocode \texttt{CAFAIL} is set when a \crd{} or \cwr{} fails. This is similar to updating a flag register.

\noindent
\textbf{Linearizability of the upgraded operations}.
The correctness follows from the fact that the \texttt{top} is read using \crd{}, which adds it to the corresponding thread's \hbrset, upon which the thread starts monitoring for any subsequent modifications to \texttt{top}.
Since the top itself is never deleted it is guaranteed to be always in the data structure at the time it is added to the \hbrset (DII). 
At the beginning of an operation the \hbrbit is clear and it is only set when the thread receives an invalidation request for the \texttt{top} when it is modified elsewhere. 
Later, the thread attempts to change the \texttt{top} using \cwr{} which atomically checks the \hbrbit for any interfering memory access. 
It fails if the \hbrbit is set.  This causes the thread to remove the top from its \hbrset, clear the \hbrbit using \utagall, and then retry the operation.
Otherwise, the thread succeeds by changing \texttt{top} to another node.
The push and pop operations can be linearized on the successful \cwr{} at \lineref{pushlp} and \lineref{poplp} in \algoref{stack}, respectively.
\begin{algorithm}[!htbp]
    \caption{Using \ca with unbounded lock free stack. Lines annotated as LP are the linearization points. 
    }
    \label{algo:stack}
    \footnotesize
    \begin{algorithmic}[1]
        \State type Node \{Key key, Node *next\}
        \State class Stack \{Node *top\}
        \State \#define \texttt{CA\_CHECK} \IfThenNoS{\texttt{CAFAIL}}{\texttt{untagAll(); goto retry;}}
        \Statex
        \Procedure{push}{key}
        	\State newtop = new node(key);
            \State retry:
            \State t $\leftarrow$ \crdp{top}; \texttt{CA\_CHECK}
            \State newtop->next = t;
            \State \cwrp{\&top}{newtop}; \texttt{CA\_CHECK} \label{lin:pushlp}\Comment{LP}
        \EndProcedure
        \Statex
        \Procedure{pop}{ }
            \State retry:
            \State t $\leftarrow$ \crdp{top}; \texttt{CA\_CHECK}
            \If {NULL == t}
                \State \texttt{untagAll(); and return;}
            \EndIf
            \State \cwrp{\&top}{t->next}; \texttt{CA\_CHECK} \label{lin:poplp}\Comment{LP}
            \State free(t) \label{lin:stackfree}
        \EndProcedure
    \end{algorithmic}
\end{algorithm}

Note, the call to free at \lineref{stackfree} is safe because
whenever a core C1 modifies the \texttt{top} (either for push or pop), any other core C2 having access to \texttt{top} will fail its \cwr{} because C1 will invalidate C2's tag by setting its \hbrbit.


\ca is ABA-safe despite the fact that it allows immediate reuse of freed objects.
Suppose a thread T1, in order to insert a new node, reads an address A from the \texttt{top} into a local variable $t$.  Then just before it executes a CAS to set \texttt{top} to $t$'s next, some other thread T2 removes A by setting a node at address B as the new \texttt{top}, frees A, and then pushes a new node at this recycled address A, making it the new \texttt{top}.
Now T1 would succeed its CAS (based on address comparison) as the \texttt{top} still contains the address A, which matches the expected address stored in its local variable $t$.  Thus it incorrectly succeeds when it should have failed\textemdash a typical ABA case.
\ca prevents this error as \cwr{}, unlike a CAS, is not based on comparing two values. Instead, it relies on the underlying cache invalidation messages to detect that a location has been modified since it was last read.




\subsection{In Data Structures having Multiple Writes with Locks}
\label{sec:ll}
Another category of linked concurrent data structures have update operations wherein threads optimistically traverses a sequence of nodes ending in multiple updates within a critical section guarded by locks. One example is the lazy list~\cite{heller2005lazy}.

Operations of data structures with such design patterns could be upgraded to enable the proposed technique's \smr using the following broad guidance:
\begin{enumerate}
    \item In the search phase, use DI to replace all reads with \crd{}. Use \utag{} to remove previously traversed nodes from current thread's \hbrset when they are no longer required to prove that a node, to be accessed in the future, is reachable in its data structure at the time it is tagged.
    If a \crd{} fails during the traversal then do \utagall{} and retry the search.
    For read only operations this will suffice. Update operations require  the following steps:
    \item Use try locks
    designed using \crd/\cwr{} (\algoref{lock}) to lock all the nodes identified at the end of the search. This marks the beginning of a critical section to execute the updates atomically. 
    If lock acquisition fails on any of the nodes then unlock the previous nodes (if any), do \utagall and retry the operation. 
    \item Within the critical section use normal writes to execute intended updates.
    This is safe because the nodes are guarded by the critical section and therefore cannot be concurrently updated or reclaimed (partial relaxation of DI). 

    \item If the update is a delete, mark the node before unlinking it. This satisfies the \emph{no unlinked access rule}. 
    
    \item Finally, unlock any locked nodes and execute \utagall before exiting the operation. Note that unlock may use regular stores instead of \cwr{},  since locked nodes cannot be freed by other threads.
\end{enumerate}

By the way of example of a lazylist (in reference to \algoref{list}) we will demonstrate how we can easily upgrade it to use \ca.

Using D1, all the regular reads are replaced by \crd{s} in searches, as shown in \algoref{list}.
As explained in the specification, the \crd{s} atomically: add a node to current thread's \hbrset, if it is not in it already, check that the \hbrbit is clear, and complete a normal read, if the condition succeeds.
However, if the \crd{} fails (when \texttt{CAFAIL} is set) then it could be the case that a subset of the nodes in the \hbrset have been modified (potentially deleted) since they were last accessed, therefore it may not be safe to read them. 
In such a case the \hbrset is emptied, the \hbrbit is unset using \utagall, and the search is retried. Otherwise it continues and eventually stops when some \pred and \curr nodes of interest are found (\lineref{locateret} in \algoref{list}).



Note, if we do not untag previous nodes during searches, then since \crd{s} tag nodes, we will have all the nodes in the search path added to a thread's \hbrset.
This could cause \crd{s} to fail when any node (relevant to current access or not) in the search path is modified, which forces operations to retry repeatedly. In other words, certain updates will be serialized, as if threads acquired a global lock, which will inhibit concurrency.
As a remedy to this problem, threads can untag previous nodes using \utag and are only required to keep two consecutive nodes tagged at any given time, which is equal to the number of nodes required to carry out updates, much like hand-over-hand locking.

Furthermore, in order to guarantee searches are safe we need to ensure that at the time a node is tagged it is reachable in the list 
(DII). To see why, assume a case where a thread accesses content of an arbitrary node using \crd{}. During this \crd{}, atomically: a cache line containing the node will be tagged and then its content will be loaded. Now, if the node was already marked before it was tagged by the \crd{}, then a subsequent \crd{} would succeed even though the node is marked (logically deleted), which is not safe as the node could be reclaimed (a \uaf error). 
This is resolved by validating that a node is not marked, immediately after the \crd that tagged it.
If the node is found to be marked, validation fails and the corresponding operation untags all nodes and retries. 
For example, in the lazy list a node is first tagged during \crd at \lineref{tagged} in \algoref{list}, due to a \func{validate()} invoked from \lineref{tagged2},
~\ref{lin:tagged3}, or~\ref{lin:tagged4}. If \func{validate()} returns False due to the node being marked then the operations untags all nodes and retries.
This way DII is satisfied.

\begin{algorithm}[t]
    \caption{\ca based lock. Precondition: the node containing the lock field should be \crd{} so that it is tagged.}
    \label{algo:lock}
    \footnotesize
    \begin{algorithmic}[1]
        \Procedure{tryLock}{bool *lock}
            \State lockVal $\leftarrow$ \crdp{lock};
            \IfThen{\texttt{CAFAIL} or 1 $==$ lockVal}{return False;}
            \Statex
            \State \cwrp{lock}{1};
            \IfThenNoS{\texttt{CAFAIL}}{return False;}
            \State return True;  
        \EndProcedure

        \Statex

        \Procedure{unlock}{bool *lock}
            \State *lock $\leftarrow$ 0; \Comment{safe as a node can only be mutated by owner.}
        \EndProcedure
    \end{algorithmic}
\end{algorithm}

\begin{algorithm}[!t]
    \caption{Using \ca with lazylist~\cite{heller2005lazy}. 
    Lines annotated as LP are the linearization points.
    }
    \label{algo:list}
    \footnotesize
    \begin{algorithmic}[1]
        \State type Node \{Key key, lock, mark, Node *next\}
        \State class Lazylist \{Node *head\}
        \State \#define \texttt{CA\_CHECK} \IfThenNoS{\texttt{CAFAIL}}{\texttt{untagAll(); goto retry;}}
        \Statex
        \Procedure{validate}{Node *node} \label{lin:validatep}
            \State isMarked $\leftarrow$ \crdp{node->mark}; \IfThenNoS{\texttt{CAFAIL}}{return False;} \label{lin:tagged}
            \IfThenElse{\texttt{isMarked}}{return False;}{return True;}
        \EndProcedure
        \Statex
        \Procedure{locate}{Key key}
            \State retry:
            \State pred $\leftarrow$ \crdp{head}; \texttt{CA\_CHECK} \label{lin:tagged1}
            \State \Call{validate}{pred}; \IfThenNoS{$False$}{\texttt{untagAll(); goto retry;}} \label{lin:tagged2}
            \State curr $\leftarrow$ \crdp{pred->next}; \texttt{CA\_CHECK} 
            \State \Call{validate}{curr}; \IfThenNoS{$False$}{\texttt{untagAll(); goto retry;}} \label{lin:tagged3}
            \State currkey $\leftarrow$ \crdp{curr->key};  \texttt{CA\_CHECK}
            \While{currkey < key}
                \State \utagp{pred};
                \State pred $\leftarrow$ \crdp{curr}; \texttt{CA\_CHECK}
                \State curr $\leftarrow$ \crdp{curr->next}; \texttt{CA\_CHECK}
                \State \Call{validate}{curr}; \IfThenNoS{$False$}{\texttt{untagAll();goto retry;}} \label{lin:tagged4}
                \State currkey $\leftarrow$ \crdp{curr->key};  \texttt{CA\_CHECK}
            \EndWhile
            \State return $\langle$ pred, curr, currkey $\rangle$; \label{lin:locateret}
        \EndProcedure
        \Statex
        \Procedure{contain}{key}
            \State $\langle$ pred, curr, currkey $\rangle$ $\leftarrow$ locate(key); \Comment{LP:When currkey was read.}
            \State \utagall{\texttt{()}};
            \State return (currkey == key);
        \EndProcedure
        \Statex        
        \Procedure{insert}{key} \label{lin:insproc}
            \State retry:
            \State $\langle$ pred, curr, currkey $\rangle$ $\leftarrow$ locate(key); \Comment{$LP$ when insert fails.}
            \IfThen{currkey == key}{\utagall{}; \& return False;}
            \If{False == tryLock(\&pred->lock)} \Comment{attempt locking pred.} \label{lin:ilockpred}
                \State \utagall{} \& retry;
            \EndIf
            \If{False == tryLock(\&curr->lock)} \Comment{attempt locking curr.} \label{lin:ilockcurr}
                \State unlock(\&pred->lock);
                \State \utagall{} \& retry;
            \EndIf
            \State node $\leftarrow$ new Node(key, curr);
            \State pred->next $\leftarrow$ node; \Comment{$LP$ when insert succeeds.} \label{lin:insadd}
            \State unlockAll() and \utagall{}; 
            \State return True;
        \EndProcedure
        \Statex
        \Procedure{delete}{key} \label{lin:delproc}
            \State retry:
            \State $\langle$ pred, curr, currkey $\rangle$ $\leftarrow$ locate(key); \Comment{$LP$ when delete fails.}
            \IfThen{currkey != key}{\utagall{}; \& return False;}
            \If{False == tryLock(\&pred->lock)} \label{lin:dlockpred} \Comment{attempt locking pred.}
                \State \utagall{}; \& retry;
            \EndIf
            \If{False == tryLock(\&curr->lock)} \label{lin:dlockcurr}\Comment{attempt locking curr.}
                \State unlock(\&pred->lock);
                \State \utagall{}; \& retry;
            \EndIf
            \State curr->mark $\leftarrow$ true; \Comment{$LP$ when delete succeeds.} \label{lin:delmark}
            \State pred->next $\leftarrow$ curr->next; 
            \State unlockAll(); \& \utagall{}; 
            \State \texttt{free(curr)};
            \State return True;
        \EndProcedure        
    \end{algorithmic}
\end{algorithm}        

One may further ask, what if the node was marked (already logically deleted) and also freed before it is tagged? In that case a subsequent \crd{} could succeed as its \hbrbit will not be set since no update will occur after the node was tagged. This could cause a \uaf error.
However, this cannot happen because, in order to free the node, a reclaimer has to unlink it by modifying the next field of its predecessor, which is already in the thread's \hbrset. Thus, if the predecessor node is modified the thread's \hbrbit will be set and the \crd will fail, preventing unsafe access. This invariant is maintained during a search that eventually yields a \pred and \curr that were reachable in list at the time they were tagged.

Later, before starting the updates, locks on the \pred and \curr nodes are acquired (\lineref{ilockpred} \&~\ref{lin:ilockcurr} for \func{insert()} and \lineref{dlockpred} \&~\ref{lin:dlockcurr} for \func{delete()}). However, it may happen that after the search returns the nodes and before the locks are acquired some thread may delete these nodes. In that case if the lock is accessed with normal reads and writes then the thread may attempt acquiring lock on a freed node which could lead to undefined behaviour: unlike \crd{s}, regular reads do not have the ability to check whether the nodes have been modified. Thus, to resolve this issue we provide \crd/\cwr{} based try locks which only acquire the lock on a node if it has not been modified (deleted) concurrently.

\algoref{lock}
depicts the implementation of this lock.
It has a precondition that the node containing the lock field should have been previously accessed using \crd{} so that it gets added to its thread's \hbrset{}, enabling a \crd/\cwr{} to verify through \hbrbit whether the node has been modified since then. 
In further detail, a thread does a \crd{} on the lock variable. If it sets \texttt{CAFAIL}, the node of which the lock is part might have been deleted; if it returns $1$, it means that lock is busy. In both the cases, the lock acquisition fails. Otherwise, a thread proceeds to acquire the lock by setting the lock field to 1 using a \cwr{}, which again checks if the node containing the lock field has not been modified (possibly deleted). 
If the check succeeds it writes 1 to the lock field and returns \texttt{True}, indicating that lock acquisition is successful. Otherwise if the \cwr{} fails (by setting \texttt{CAFAIL}) it returns \texttt{False} indicating that lock acquisition has failed, and the operation which invoked the lock untags all nodes and retries.

The insert operation(\lineref{insproc}, \algoref{list}), first executes \texttt{locate}, which returns tagged \pred and  \curr nodes along with \texttt{currkey} (key field of \curr). If the key to be inserted is already present in the list then the operation returns false. Otherwise, the key is not present and needs to be inserted.
To insert the key, first the \ca based trylocks on the \pred and \curr nodes are acquired, then a new node is created and inserted between pred and curr. Following that all locks are released, nodes are untagged, and then the operation returns true.

Note, because the \pred and \curr nodes are already locked, no other thread could ever modify them without acquiring lock first. Therefore, validation to check whether the node has been concurrently freed is not needed. This allows us to use normal reads and writes instead of \crd/\cwr{} within the critical section. 

The delete operation(\lineref{delproc}, \algoref{list}) invokes \texttt{locate}, which returns tagged \pred and \curr nodes along with currkey. If the key to be deleted is not present in the list then the operation untags the nodes and returns false.
Otherwise, similar to the insert operation, it acquires the trylocks on both nodes, does a write on the \curr node to set its \texttt{mark} field, unlinks the \curr node, unlocks and untags both the nodes, frees the \curr node and then returns true.

\subsubsection{Correctness}
If all the aforementioned rules are followed to enable \ca in the lazylist then the list is linearizable and all access in it are safe. \texttt{Contains} or unsuccessful \texttt{inserts} and \texttt{deletes}, which behave like \texttt{contains}, can be linearized at the time when the key of \curr was read. Whereas, successful \texttt{inserts} and \texttt{deletes} can be linearized when a new node is linked (\lineref{insadd}) or a node is marked (\lineref{delmark}), respectively. Also, because \crd{s} never dereference an unreachable (unlinked) node, \uaf errors do not occur. Therefore the lazy list with \ca is safe. The following section discusses the correctness in detail.

\begin{assumption}
List operations are implemented using the provided guidelines to enable \ca based immediate reclamation.
\begin{enumerate}
    \item Replace and Analyse (DI) rule is followed. 
    \item Validate Reachability (DII) is followed. In order to ensure that only reachable and unmarked nodes are tagged, marked field of a node is validated after it is tagged.
    \item In updates, normal locks are replaced with \ca based try locks.
    \item Before freeing a thread must write on the node (i.e update the mark field).
\end{enumerate}
\label{asm:wellbehave}
\end{assumption}

For \ca based programs to be correct, all data structure operations should do shared memory reads and writes using \crd{} and \cwr{}, a thread is required to write on a node before freeing it, and when a node is tagged it should be reachable and unmarked. 
Additionally, for \textbf{DI} to work we assume that the underlying hardware provides cache coherency mechanism like MSI, MESI or other such equivalent mechanisms. 


\begin{property}
\label{prop:cAccessAtomic}
A node is tagged the first time its content is \crd{} and it stays tagged until explicitly untagged using \utag{} or \utagall.
\end{property}

\begin{property}
\label{prop:writeinvalidates}
When any field of a node is written the underlying cache coherency mechanism invalidates all the cached copies of the location at other threads by setting their \hbrbit to true, which stays true unless unset by \utagall.
\end{property}


\begin{claim}
\label{clm:cAccessSucc}
A conditional access i.e. a \crd or \cwr on a node only succeeds if none of the tagged nodes have been invalidated since they were tagged.
\end{claim}

\begin{lemma}
\label{lem:correcttag}
A node's predecessor is unmarked and its next field points to the node at the time it is tagged.
\end{lemma}
This implies that a thread will never \crd{} next field (or discover new nodes) from a marked node. 
\begin{proof}
We prove the lemma by induction on the sequence of nodes in the \ca based lazylist. \\
\noindent
\textbf{Base case:} Initially when the search starts it does a \crd on \head, thus loads the address of head node in a local pointer \pred. 
Subsequently, the head node's mark field is \crd{} which atomically: tags the head node and loads the value of its mark field. Since the head pointer is never changed and the head node is never marked both the head node is valid when it was tagged. And any subsequent \crd{} on the fields of head node would fail if any of the fields of the head node change. 
Later, the head node's next field is \crd{} and saved in a local pointer \curr followed a \crd{} on the mark field of the \curr node. The latter \crd{} tags the curr node and then loads the value of its mark field.

So at this point, when the \curr node is tagged its predecessor \pred is guaranteed to be not marked because a head node is never marked and its next field is guaranteed to be unchanged (still points to \curr) because otherwise the \crd on \curr nodes's mark field would have failed by virtue of \propref{cAccessAtomic}, \propref{writeinvalidates}, and \clmref{cAccessSucc}. Also note that the \curr node's mark field is verified to be unmarked. So, both the \pred and \curr nodes are unmarked and \pred points to \curr for the base case.
\\
\noindent
\textbf{Induction Hypothesis:} Say, this is true for any arbitrary \pred and \curr nodes in search path of a thread. That is both the \pred and \curr nodes are unmarked and \pred points to \curr. Now we shall prove that at the time a node next to \curr, say \suc, is tagged then the \curr is still unmarked and points to \suc. \\
\noindent
\textbf{Induction Step:}  
In order to be able to access \suc to tag it, a thread first does a \crd{} on its predecessor node that is \curr to load its next field. If the \crd{} succeeds implies that \curr node hasn't changed since it was tagged. Meaning neither it has been marked or its next field has changed. Therefore, when marked field of \suc is \crd to be tagged its predecessor is unmarked and points to \suc. Otherwise, the \crd would fail by virtue of the \propref{cAccessAtomic}, \propref{writeinvalidates}, and \clmref{cAccessSucc}. 

Hence, the lemma is true for an arbitrary node.
\end{proof}

\begin{theorem}[\ca is Safe]
\label{thm:hbrsafe}
Threads cannot access reclaimed nodes in \ca based list.
\end{theorem}
\begin{proof}
Let us assume the contrary. That is thread could access reclaimed nodes. This may happen in \ca based lazylist in following two scenarios:
\begin{enumerate}
    \item A thread reclaims a node without setting its mark field to true. In this case, readers which have the node tagged will not receive an invalidation message and thus will not have their \hbrbit set. As a result the readers, unaware of the deletion, will succeed subsequent \crd{} which could cause \uaf error. But this is not possible because we assume that reclaimer's rule is followed.
    \item A thread tags a node after it has been marked so that subsequent \crd{} or \cwr{} will not receive any invalidation request for the node. And, therefore will not have their \hbrbit set. Which in turn would cause conditional accesses to succeed. This case too is not possible because of the \lemref{correcttag}.  
\end{enumerate}
Hence, both the the scenarios contradict our assumption implying use of \ca in lazylist is safe.
\end{proof}







\begin{theorem}[\ca is ABA free] Conditional accesses are immune to ABA problem.
\label{thm:abasafe}
\end{theorem}
\begin{proof}
Assume, that the \ca based lazylist have ABA problem. Therefore, there exists a case where a memory location \textit{m} is conditionally accessed by a thread $T_i$ (A), and then \textit{m} is modified by another thread $T_j$ (B), followed by another conditional memory access on \textit{m} by $T_i$ (A), such that the second memory access succeeds without recognising the fact that \textit{m} was changed since its first access. 

Since, whenever a memory location is first \crd{} it is tagged, $T_i$ tags \textit{m} when it does its first \crd{} on \textit{m}. Later if \textit{m} is modified by $T_j$ then by virtue of the cache coherence mechanism it invalidates all the remote copies of \textit{m}, in process invalidating the tags (\propref{writeinvalidates}). 
Therefore, when $T_i$ does \crd{} again on \textit{m} it will find that \textit{m} has been invalidated causing it to fail (\clmref{cAccessSucc}). Since, a \crd{} cannot fail and succeed at the same time, our initial assumption that the second \crd{} on \textit{m} by $T_i$ succeeds even when another thread modifies it after $T_i$ first accessed it is wrong. Therefore, ABA issue cannot occur.

\end{proof}

\subsubsection{Progress}

If \ca instructions are implemented in hardware such that spurious failures due to associativity evictions or interrupts are eliminated and conditional accesses could only fail due to real data conflicts then searches for \ca based data structures could be guaranteed to be lock-free. Because, a failure of a conditional access would imply that some thread successfully executed an update on a conflicting address and thus at least one thread is guaranteed to make progress.

On the other hand, if the implementation of proposed \ca semantics could not guarantee freedom from spurious failures then a fallback technique could be used. 
\\
\\
\noindent
\textbf{facilitating progress}
As noted conditional accesses are vulnerable to spurious failures due to limits on hardware resources, like overflow of tagged cachelines because of associativity evictions. To reduce the probabilty of such spurious failures we propose \utag{} which helps in maintaining only a minimal set of tagged references by untagging previous locations which will not be required to establish correctness of searches.

\section{Experiments}
\label{sec:exp}

We prototype \ca (CA) using the Graphite multicore simulator~\cite{miller2010graphite}.  Our modifications were restricted to the L1 data cache level; we did not change the cache coherence protocol.
Graphite is configured to use a directory based MSI cache coherency protocol with a private 32K L1 and a shared inclusive 256K L2 cache. Each cacheline is 64 bytes and each thread runs on a dedicated simulated core with a basic branch prediction mechanism and an out-of-order memory subsystem.

We evaluate the scalability and memory efficiency of CA using microbenchmarks that stress test the lazy list and an external binary search tree (extbst).  We also use stack and hash table microbenchmarks to evaluate CA at different contention levels.
The keys in the lazy list, stack and hash table range from 0 to 1K; the extbst keys range from 0 to 10K. The hash table has 128 buckets,
where each bucket is a lazy list.  

Each of these data structures are made to use the following safe memory reclamation techniques: a leaky implementation(\textbf{none}:), \ca (\textbf{ca}), the 2geibr variant of IBR (\textbf{ibr}), \textbf{rcu}, quiescent state based reclamation (\textbf{qsbr}), hazard pointers(\textbf{hp}), and hazard eras (\textbf{he}). CA reclaims each deleted node immediately and requires no other parameters. The other reclamation schemes were configured to attempt reclamation after every 30 successful remove operations (\textit{reclamation frequency}). For epoch based schemes (ibr, rcu, qsbr and he) the epoch were configured to change after every 150 allocations (\textit{epoch frequency}). These values are the default in the IBR benchmark\cite{wen2018interval}.

Each trial in each experiment prefills its data structure to 50\% full and executes 3K operations per thread. The number of threads varies from 1 to 32. Each time a thread invokes a data structure operation, it randomly chooses an operation with a random key. In our experiments threads choose insert or delete with equal probability of 0\%, 5\% or 50\%, allowing us to run experiments with 0\% (read only), 10\% and 100\% updates, respectively.
Because the insert and delete probabilities are equal in all our workloads the data structure size remains roughly constant, storing half the elements in the key range.
For each workload configuration we report the average of three runs. There was no significant variance across the runs.

\begin{figure*}[h]
     \begin{minipage}{\textwidth}
        \begin{subfigure}{\textwidth}
            \includegraphics[width=0.33\linewidth, height=6cm, keepaspectratio]{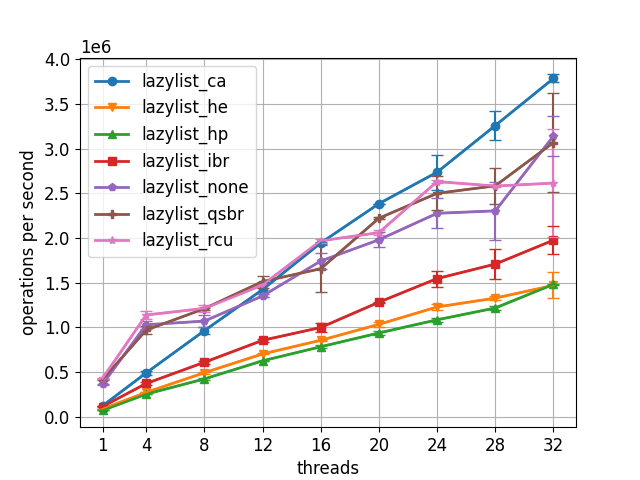}\hfill
            \includegraphics[width=0.33\linewidth, height=6cm, keepaspectratio]{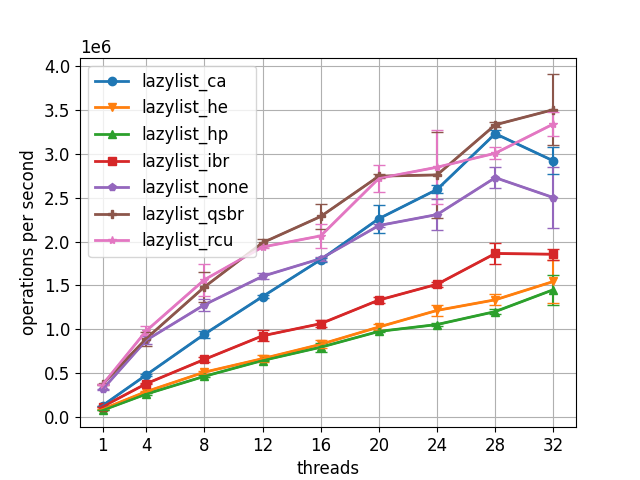}\hfill
            \includegraphics[width=0.33\linewidth, height=6cm, keepaspectratio]{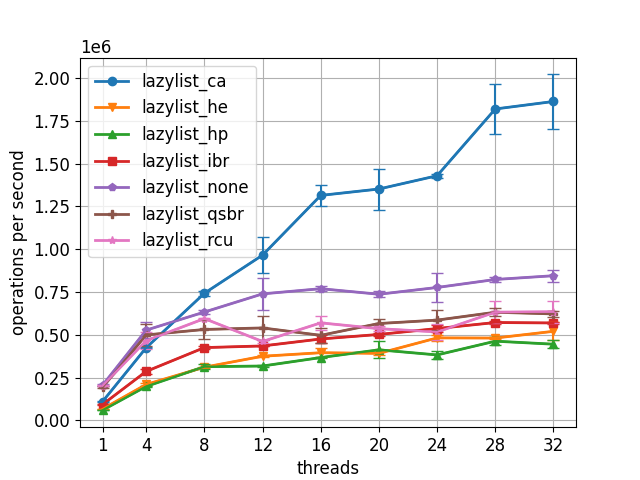}\hfill
            \label{fig:list}
        \end{subfigure}
        \begin{subfigure}{\textwidth}
            \includegraphics[width=0.33\linewidth, height=6cm, keepaspectratio]{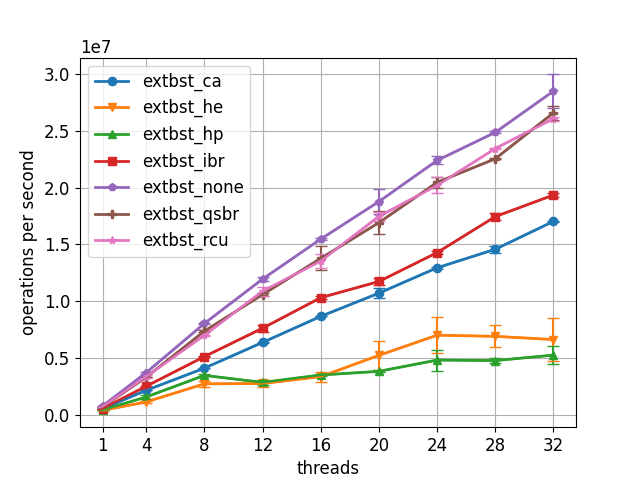}\hfill
            \includegraphics[width=0.33\linewidth, height=6cm, keepaspectratio]{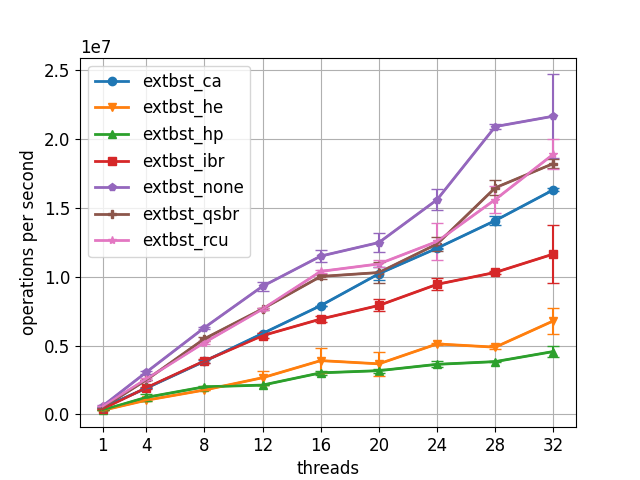}\hfill
            \includegraphics[width=0.33\linewidth, height=6cm, keepaspectratio]{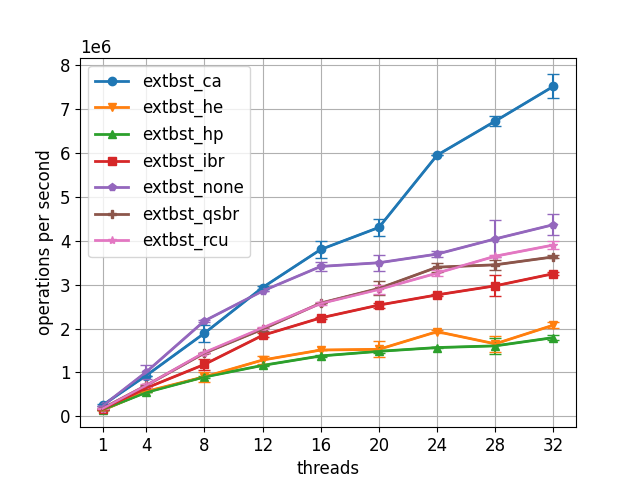}\hfill
            \label{fig:extbst}
        \end{subfigure}        
     \end{minipage}

    \caption{Evaluation of throughput. Y axis: throughput. X axis: \#threads. Left: 0i-0d. Middle: 5i-5d. Right: 50i-50d. (Top Row) Lazy linked-list, size:1K. (Bottom Row) External BST, size:10K. 
    }
    \label{fig:exp1}
\end{figure*}

\begin{figure*}[h]
     \begin{minipage}{\textwidth}
        \begin{subfigure}{\textwidth}
            \includegraphics[width=0.33\linewidth, height=6cm, keepaspectratio]{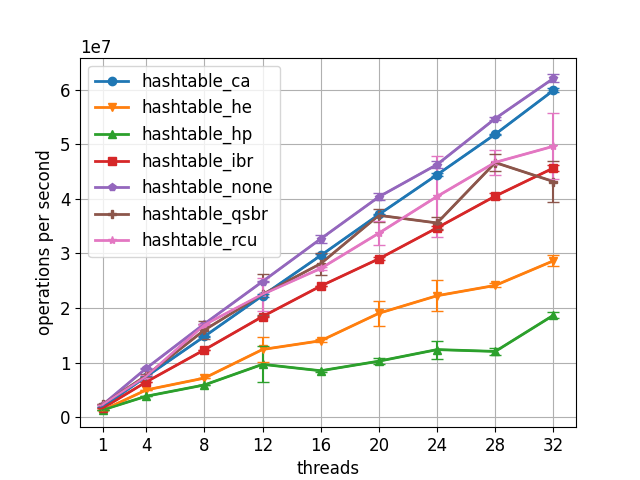}\hfill
            \includegraphics[width=0.33\linewidth, height=6cm, keepaspectratio]{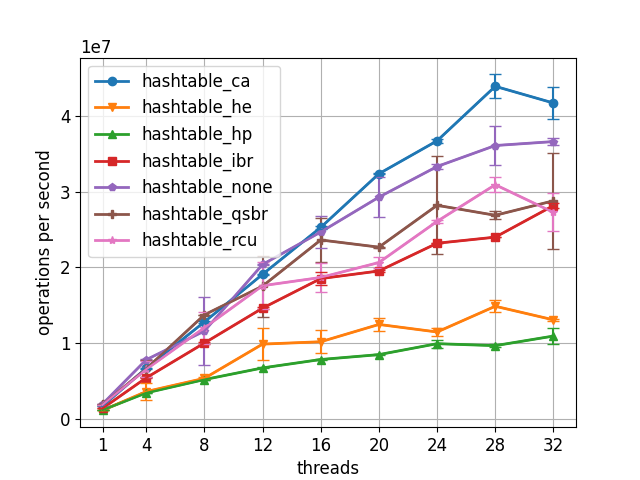}\hfill
            \includegraphics[width=0.33\linewidth, height=6cm, keepaspectratio]{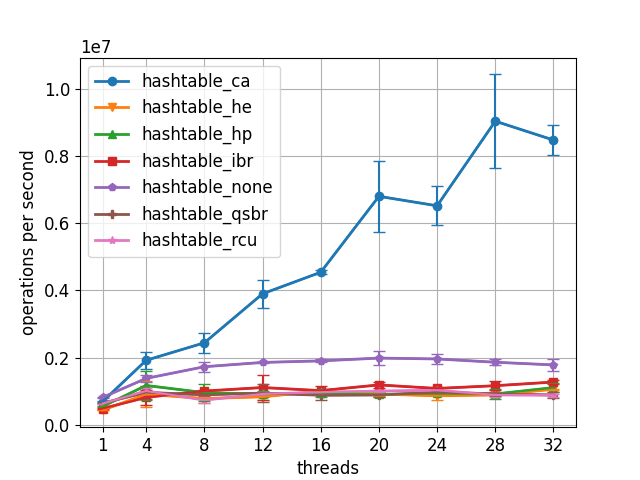}\hfill
            \label{fig:hashtable}
        \end{subfigure}
        \begin{subfigure}{\textwidth}
            \includegraphics[width=0.33\linewidth, height=6cm, keepaspectratio]{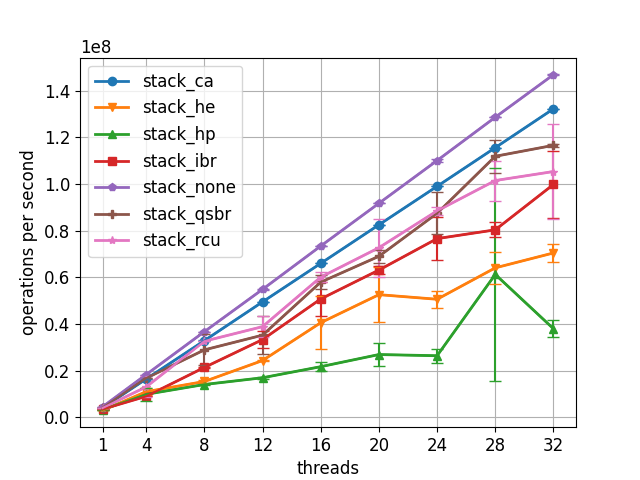}\hfill
            \includegraphics[width=0.33\linewidth, height=6cm, keepaspectratio]{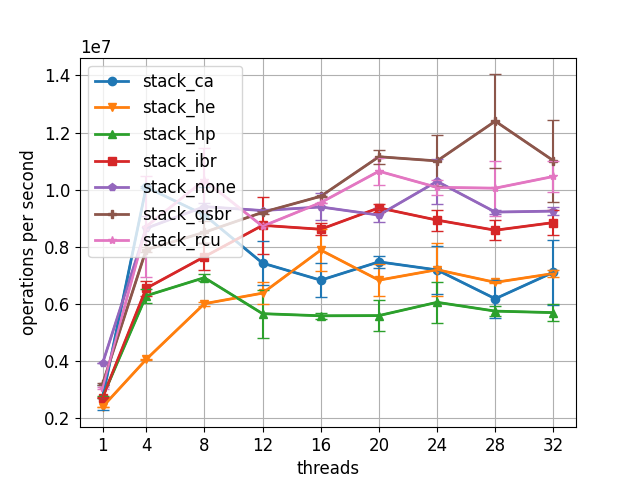}\hfill
            \includegraphics[width=0.33\linewidth, height=6cm, keepaspectratio]{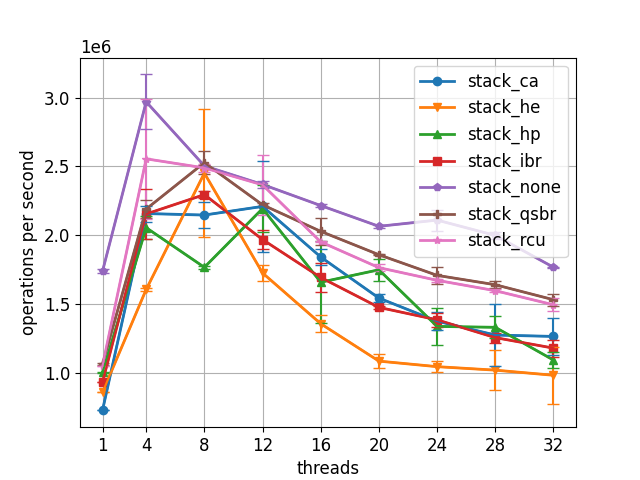}\hfill
            \label{fig:stack}
        \end{subfigure}
     \end{minipage}
    \caption{Evaluation of throughput. Y axis: throughput. X axis: \#threads. Left: 0i-0d. Middle: 5i-5d. Right: 50i-50d. (Top Row) Chaining Hash table, \#buckets:128, MAX size 128K. (Bottom Row) Stack.
    }
    \label{fig:exp2}
\end{figure*}

Throughout the experiments in \figref{exp1} and \figref{exp2}; \texttt{hp}, \texttt{he} and \texttt{ibr} are generally slower than the other algorithms.
This mainly can be attributed to high per-read overheads, as these algorithms have read/write fences to access or update reservations and epochs, respectively. 
Additionally, these algorithms have reclamation overhead which requires scanning of reservations to determine which records are safe to free.
In general this results in poor cache behaviour and high operation latency.

On the other side, \texttt{rcu} and \texttt{qsbr} have no per-read overhead.  Their main overhead arises from their reclamation events, where batches of retired objects are freed after scanning the epochs of all the processes.  This is amortized over multiple operations. As a result these algorithms are faster and perform similar to the baseline \texttt{none}, across workloads and data structures. 

In read-only workloads, CA is comparatively slower than \texttt{rcu}, \texttt{qsbr} and \texttt{none}.  This is due to the increased latency: checking the \hbrbit after each \crd{} increases the instruction count.  Since there are no conflicts, these checks are superfluous.
However, in workloads with updates, CA is closer to or faster than \texttt{rcu}, \texttt{qsbr} and \texttt{none}. 
It even outperforms these algorithms in high contention scenarios (i.e., high updates and high thread counts). This is due to the fact that CA avoids read-write fences for both readers and reclaimers. 
CA brings additional benefits.  Immediate reclamation improves cache and TLB locality, especially relative to \texttt{none}; it discovers failures earlier than other algorithms, which enables it to restart without wasting as much work; and it avoids some cache miss latencies.
All these contribute to low latency and higher throughput.  The low cost of cache misses is due to a property that unlike regular reads, in \crd{s} the impact of cache misses remains confined to its core \cite{alistarh2020memory}. We explain this in following paragraphs.

In data structure operations with normal reads and writes, all threads that share memory locations experience latency due to cache misses. Consider a lazy list, and suppose that thread T1 is about to acquire locks on its \pred and \curr nodes.  Suppose that another thread T2 has read the \pred and is about to re-read it.
At this point, at the cache level, both T1 and T2 will have copies of the cache lines corresponding to these addresses in the shared state.
When T1 acquires a lock on \pred it does a write.  This triggers coherence traffic: all other readers of \pred must invalidate their copies of the cache line (here T2).  When T2 reads \pred again it will suffer a cache miss as its copy of the cache line is invalid. In order to serve the cache miss:
\begin{itemize}
    \item T2 triggers a cache level transaction to fetch the latest copy of the cache line and waits for a response.
    \item T1, which has the cache line in M state may be forced to write its copy of the cache line back to the memory hierarchy, and also supply it to T2. 
\end{itemize}

This wastes T2's compute time, because T2 will ultimately see that the line has changed, necessitating that it restart its operation.  If it had not waited, it could have already restarted and executed multiple instructions. 
Furthermore, since T1 acquired the lock on \pred it is likely to write to \pred again.  T2's request caused the line to downgrade from M to S in T1's cache, so a subsequent write by T1 will need to begin with an ownership request that causes T2 to re-invalidate the line. Such frequent downgrading to shared state and upgrading to modified state interferes with the gains made by write buffering and makes it difficult to hide the cache latency. These overheads worsen with increases in contention on shared locations.

On the other hand, in data structures designed using \crd{} and \cwr{}, T2's second \crd{} will fail validation and retry its operation by detecting that the line is no longer present, \emph{without requesting a new copy of the line}.  Unlike the aforementioned issues with regular reads/writes, CA allows T2 to skip requesting the value of \pred, which prevents global cache traffic.  This avoids read latency for T2, and also helps T1 to avoid a cache state upgrade transactions. In other words, unlike regular read/write based data structures, the impact of failure to access cache lines in data structures with CA remains confined to a local core~\cite{alistarh2020memory}. 

Thus, in all the data structure implementations, the lower L1 data cache latencies that result from the aforementioned properties allow CA to be as good as the other algorithms, if not faster, when contention is high.  In read only workloads, CA is slower or comparable to the baseline (none) and other fast algorithms (\texttt{qsbr} and \texttt{rcu}) mainly due to overhead of its higher instruction count. 

\begin{figure}[h]
\centering
\includegraphics[width=0.5\textwidth]{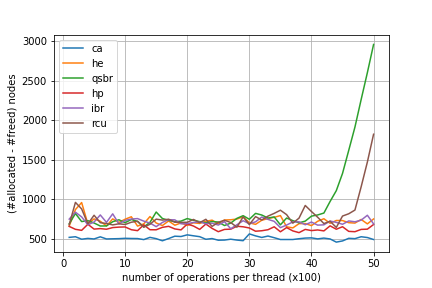}
\caption{Memory Consumption: Shows Number of nodes allocated but not yet freed for a list of size $\sim$500. Insert and delete percent is 50 each. For 16 threads.}
\label{fig:memconsume}
\end{figure}

\figref{memconsume} looks at memory overheads: For each of the reclamation schemes, we measure the number of nodes that were allocated but not yet freed (Y axis) during execution of the lazy list data structure after every 1000 operations (X axis).  This test exposes the amount by which the memory footprint of a data structure increases when paired with different reclamation schemes. For this experiment, we use a lazy list with values in the range of 0 to 1000, initially pre-filled with 500 nodes.  The experiment has 16 threads operating on the list.  During the measured part of the experiment all threads execute insert and delete operations with equal probability of 50\% (100\% update workload). Each thread runs 5000 ops. 

In the ideal case, at any time during the experiment the list size should be roughly 500 due to the workload characteristics, and the number of nodes deleted but yet not freed should be zero.  The CA scheme has a consistent reading of roughly 500 nodes that are allocated but not freed; these are the nodes that are still reachable in the list. This confirms that we are achieving immediate reclamation and keeping the memory footprint low.  Since the other reclamation schemes defer memory reclamation of deleted nodes by collecting them in a local retired list, the number of nodes that have not been reclaimed increases, which in turn leads to increased memory footprint.  This intuition is verified in the chart, as hp, he, ibr, rcu and qsbr all report a higher number of un-reclaimed nodes It is worth mentioning that, since in qsbr and rcu a delayed thread could prevent reclamation of all threads, the number of unreclaimed nodes could increase without bound.  Had we run the experiment for longer, the number of unreclaimed nodes for these schemes would be expected to balloon as soon as any thread context switched.
\section{Related Work} 
\label{sec:related}
\subsection{Discussion on Reclamation Techniques}
Many existing safe memory reclamation techniques delay reclamation of unlinked nodes, which could be broadly categorised as epoch based reclamation (EBR), hazard based reclamation (HBR), reference counting based reclamation (RCBR) and hybrid reclamation (HYR): using a combination of prior techniques or specialized hardware support.
EBR~\cite{hart2007performance,fraser2004practical,mckenney1998read, brown2015reclaiming} schemes are fast but could have an unbounded number of unreclaimed nodes. HBR~\cite{michael2004hazard, herlihy2005nonblocking, dice2016fast} and RCBR~\cite{detlefs2002lock, herlihy2005nonblocking, blelloch2020concurrent, nikolaev2019hyaline} can bound the number of unreclaimed nodes but are generally slow due to high per-read overhead or node instrumentation overhead. HYR techniques have achieved both speed and bounds on unreclaimed number of nodes with varying success, but require assumptions pertaining to specialised hardware or operating system and memory allocators~\cite{nikolaev2019hyaline, ramalhete2017brief, wen2018interval, nikolaev2020universal, singh21nbr, alistarh2017forkscan,alistarh2018threadscan, alistarh2014stacktrack,dragojevic2011power, balmau2016fast, cohen2015efficient,cohen2015automatic, cohen2018every, nikolaev2021snapshot, sheffi2021vbr, nikolaev2021crystalline, correia2021orcgc, anderson2021concurrent, ben2021space}. Nevertheless, these techniques still prefer to reduce the reclamation algorithm's overhead by delaying reclamation using batches that increase the memory footprint. Surveys of these batch-based reclamation techniques appear in~\cite{brown2015reclaiming, singh21nbr, sheffi2021vbr}. In this section we focus on techniques which could provide immediate reclamation~\cite{zhou2017hand, sheffi2021vbr} and therefore are most closely related to \ca.

Zhou et al.\cite{zhou2017hand} make use of a sequence of short hardware transactions which execute in hand over hand fashion to design concurrent data structures that retain the property of immediate memory reclamation. 
The technique relies on augmenting the data structure with a table of metadata, which can be a source of false conflicts.  Consequently, it does not appear to be as general as \ca.  Moreover, we found that the frequent starting and committing of transactions for read-only operations introduced significant latency.


VBR~\cite{sheffi2021vbr} attaches metadata to each mutable field of each node in a concurrent data structure.  
It also requires a type preserving allocator, where unlinked nodes can never be returned to the operating system.
Threads can detect use-after-free errors through the per-field metadata, which is updated atomically with the corresponding field.
While VBR can support immediate reclamation, it is most efficient when it waits until it has a batch of nodes to reclaim in a single operation.

On the other hand, \ca does not require any metadata to achieve safe reclamation and only makes use of the implicit book-keeping of the underlying cache-coherence protocol. 
In addition, since it does not make any assumptions about the number of threads present in the system, it is fully adaptive~\cite{herlihy2003space}. Furthermore, whereas HTM can accelerate timing-based attacks by leveraging the immediacy with which a thread is aborted upon a memory conflict~\cite{Lipp2018meltdown}, in particular, transaction rollbacks could lead to data leaks~\cite{intelTAA}, we believe \ca is less risky, since threads must poll to learn of remote coherence events.

\subsection{Discussion on Similar Synchronization Techniques}

\ca is inspired by, but quite different from, the Memory Tagging proposal of Alistarh et al.~\cite{alistarh2020memory}. Perhaps the most significant difference is that \ca solves the safe memory reclamation problem (and moreover offers immediate reclamation), in addition to providing useful synchronization primitives for designing concurrent data structures. In contrast, Memory Tagging does not address the memory reclamation problem, and it requires a data structure designer to rely on separate safe memory reclamation algorithms, which come with their own tradeoffs.

In reference to the programming interface, \ca offers \crd, which is critical to our immediate memory reclamation technique. The \crd instruction has no equivalent instruction in Memory Tagging, and it is not clear how one could implement \crd using memory tagging.
We also streamlined tagging by integrating it into \crd, whereas Memory Tagging requires a programmer to use an explicit \textit{AddTag} instruction before reading.

From an implementation standpoint, \ca does not require changing the underlying coherence protocol, whereas Memory Tagging’s Invalidate and Swap (IAS) instruction does, as this single instruction can invalidate many (potentially non-contiguous) remote cache lines (potentially spanning many pages). Moreover, \ca requires only 1 bit per cache line (2 bits per cache line in case of 2-way hyperthreading), whereas Memory Tagging needs to additionally maintain a set of addresses to invalidate with IAS.

It is worth noting that at the outset \ca may appear similar to HTM with early release (ER)~\cite{sonmez2007unreadtvar, skare2006early}. Possibly, one could achieve many aspects of our work by using HTM with early release. However, this would introduce various downsides.
HTM defaults to putting all reads and writes into the read/write sets. This includes the stack, the allocator, library code, etc. In data structures, many reads and writes would need to be released, which would increase the instruction count significantly. This could reduce performance and might yield a less convenient interface than \ca.

Practically, some commercial HTMs have a region based (not per access) \textit{disable tracking} feature, for instance, Intel's new TSXLDTRK~\cite{intelSDM}, and IBM's TSUSPEND/TRESUME~\cite{le2015transactional}, but this does not \textit{release} load tracking of already-read locations, it only prevents tracking of future accesses. This is different from \ca's proposed \texttt{untag} instruction which allows the release of any previously accessed location. Among Early Release proposals, we are not aware of any that release writes, although AMD's 2008 ASF proposal allowed per-access decisions about whether or not to track~\cite{christie2010evaluation}. However, ASF remains unimplemented. Additionally, we have not experimented with TSXLDTRK, but TSUSPEND suffers from relatively high overhead.

Unlike HTM, \ca does not need a write set at all, which admits a simpler implementation in hardware, as well as simpler conflict tracking and resolution.
We think \ca solves an important problem for optimistic data structures with less hardware (as demonstrated in \secref{hwimpl}). Our hope is that our hardware-software codesign approach to \ca will enable the concurrent data structure community to discover novel and efficient solutions to existing concurrency problems.

\section{Conclusion}
\label{sec:conclusion}

In this paper, we introduced \ca, a hardware extension that enables concurrent data structures to reclaim memory immediately, without introducing new inter-thread coordination.
\ca is fast.
Unlike its competitors, it does not require tuning to achieve high performance, and is tailored to the needs of modern optimistic data structures. 

To date, we have used \ca for simple nonblocking data structures, as well as optimistic lock-based data structures.
In the future, it would be interesting to determine whether \ca\ can also be used for more complex lock-free data structures.
We also believe that there are exciting opportunities at the interface between \ca and non-volatile main memory technologies.

\section*{Acknowledgments}
This work was supported by: the Natural Sciences and Engineering Research Council of Canada (NSERC) Discovery Program grant: RGPIN-2019-04227, the Canada Foundation for Innovation John R. Evans Leaders Fund with equal support from the Ontario Research Fund CFI Leaders Opportunity Fund: 38512, NSERC Discovery Launch Supplement: DGECR-2019-00048, National Science Foundation under Grant No. CNS-CSR-1814974, and the University of Waterloo. The findings and opinions expressed in this paper are those of the authors and do not necessarily reflect the views of the funding agencies.
We also thank anonymous reviewers for helping us improve the manuscript.

\bibliographystyle{IEEEtran}
\bibliography{references}

\end{document}